\documentclass[usenatbib]{mn2e}
\usepackage{amsmath, amssymb}
\usepackage{amsfonts}
\usepackage{epsfig,rotating}

\def\simeq{
\mathrel{\raise.3ex\hbox{$\sim$}\mkern-14mu\lower0.4ex\hbox{$-$}}
}

\def\ltsima{$\; \buildrel < \over \sim \;$}
\def\simlt{\lower.5ex\hbox{\ltsima}}
\def\gtsima{$\; \buildrel > \over \sim \;$}
\def\simgt{\lower.5ex\hbox{\gtsima}}

\def\msun{{\rm M_{\odot}}}

\def\be{\begin{equation}}
\def\ee{\end{equation}}

\def\del#1{{}}
\def\ltsima{$\; \buildrel < \over \sim \;$}
\def\simlt{\lower.5ex\hbox{\ltsima}}
\def\gtsima{$\; \buildrel > \over \sim \;$}
\def\simgt{\lower.5ex\hbox{\gtsima}}

\newcommand{\apj}{ApJ}
\newcommand{\apjs}{ApJS}
\newcommand{\mnras}{MNRAS}
\newcommand{\aap}{A\&A}
\newcommand{\araa}{ARA\&A}
\newcommand{\apjl}{ApJL}

\newcommand{\nat}{Nature}
\newcommand{\nar}{NewAR}

\title[AGN must be very efficient at powering outflows]{AGN must be very efficient at powering outflows}
\author[Kastytis Zubovas]{Kastytis Zubovas$^{1,2,\star}$ \\
  $^{1}$Center for Physical Sciences and Technology, Saul\.{e}tekio av. 3, Vilnius LT-10257, Lithuania \\
  $^{2}$Astronomical Observatory, Vilnius University, Saul\.{e}tekio av. 3, Vilnius LT-10257, Lithuania\\
  $^{\star}$ {E-mail:~} {\rm kastytis.zubovas@ftmc.lt} }

\begin{document}

\maketitle

\begin{abstract}

  Galaxy evolution is affected by competing feedback processes.
  Stellar feedback dominates in low-mass galaxies, while AGN feedback
  predominantly affects massive ones. Recent observational results
  reveal the dependence of black hole accretion rate (BHAR) and star
  formation rate (SFR) on galaxy stellar mass, and give information on
  the galaxy mass at which the changeover between dominant feedback
  mechanisms occurs. I use this information to derive an empirical
  estimate of the coupling efficiency, $f_{\rm AGN}$, between AGN
  luminous energy output and AGN-driven galactic outflows, and the
  momentum loading factor $f_{\rm p,AGN}$ between the momentum of AGN
  radiation field and the outflow. The results are independent of any
  particular model of AGN feedback and show that AGN feedback must be
  very efficient and/or have very large momentum loading in order to
  explain current observations. I discuss possible ways of reaching
  the required efficiency and loading factor, and the selection
  effects that might result in only weak outflows being observed,
  while the most powerful ones may be generally obscured. There are
  significant uncertainties involved in the derivation of the result;
  I suggest ways of reducing them. In the near future, better
  estimates of coupling efficiency can help distinguish among AGN
  feedback models, investigate the redshift evolution and mass
  dependence of feedback efficiency.
  
\end{abstract}


\section{Introduction}

Stellar and active galactic nuclei (AGN) feedback is a crucial
component of galaxy evolution \citep{Shankar2006ApJ,
  Kormendy2013ARA&A, Somerville2015ARA&A}. Cosmological simulations
show that the present-day galaxy mass function and its redshift
evolution can only be explained when the effects of feedback are
included \citep{Volgelsberger2014MNRAS, Schaye2015MNRAS}. The two
sources of feedback are important over different mass ranges; in
particular, stellar feedback regulates the processes in low-mass
galaxies while large galaxies are mainly regulated by AGN feedback
\citep{Shankar2006ApJ, Kormendy2009ApJS, Schaye2010MNRAS,
  Heckman2014ARA&A}.

The galaxy stellar mass at which the two processes become comparably
important can inform our understanding of the efficiency with which
feedback processes operate. This mass has been usually associated with
the `break' seen in the galaxy mass function \citep[$M_{\rm br} \sim
  1.2 \times 10^{10} \msun$;][]{Shankar2006ApJ} or, equivalently, the
maximum of the stellar-to-halo mass ratio \citep[$M_{\rm br} \sim 2
  \times 10^{10} \msun$;][]{Behroozi2013ApJ}. Recently, a similar
break was discovered in the $M_{\rm BH}-\sigma$ relation between
supermassive black hole (SMBH) mass and host galaxy spheroid velocity
dispersion, which corresponds to a transition mass $M_{\rm tr} =
3.4\pm2.1 \times 10^{10} \; \msun$ \citep{Martin2018arXiv}.

A straightforward interpretation of this break, or transition, mass is
that in galaxies with lower stellar mass, stellar feedback injects
more energy than AGN feedback into the interstellar medium (ISM) over
the lifetime of the galaxy, while in more massive galaxies, the
opposite is true. In more massive galaxies, therefore, galaxy
properties are more tightly correlated with properties of the SMBH,
while in less massive galaxies, this correlation is weaker and SMBH
mass depends less strongly on galaxy mass. Both stellar and AGN
feedback power are proportional to the corresponding mass flow rate
$\dot{M}$: the star formation rate (SFR) for stellar feedback and the
black hole accretion rate (BHAR) for AGN. Recent discoveries that the
average BHAR-to-SFR ratio increases with increasing galaxy stellar
mass \citep{Yang2017ApJ, Yang2018MNRAS} suggests a simple explanation
that even if stellar and AGN feedback processes act identically in
galaxies across the mass range, AGN feedback can become dominant for
galaxies with large stellar mass.

This interpretation offers a possibility to constrain the efficiencies
of stellar or AGN feedback coupling, i.e. the fractions of stellar or
AGN luminous energy output that is transferred to the ISM of the host
galaxy. Stellar feedback has been investigated in significant detail
\citep[e.g.,][]{Leitherer1992ApJ, Thornton1998ApJ, Murray2005ApJ,
  Walch2015MNRAS}, leading to a good understanding of the energetics
of the process. AGN feedback, on the other hand, is less well
understood, with models producing different predictions of coupling
efficiency \citep[e.g.,][]{King2003ApJ, Sazonov2005MNRAS,
  Murray2005ApJ, Zubovas2012ApJ, Ishibashi2018MNRAS} and observations
calling various theoretical scenarios into question.

In this paper, I derive a model-independent constraint on the
coupling efficiency between the AGN luminous energy output and the gas
in the host galaxy, $f_{\rm AGN}$, averaged over the Hubble time. This
efficiency turns out to be very high, $f_{\rm AGN} > 0.045$. This
implies that most galaxy quenching via AGN outflows happened at high
redshift and that most massive galactic outflows seen today are far
less efficient than they should have been in high redshift galaxies,
or that stellar feedback is much less efficient than generally
assumed. I suggest ways of testing these findings and discuss their
importance in distinguishing among AGN feedback models.

\section{AGN feedback coupling efficiency} \label{sec:couple}

In order to compare the effects of stellar feedback and AGN feedback,
I consider the injection of energy and/or momentum by these
processes into the ISM integrated over the lifetime of the
galaxy. This approach is also advantageous, because it doesn't require
any detailed analysis of the star formation or accretion histories of
the galaxy, but merely the time-integrated or averaged values of
$\dot{M}$. Furthermore, injection of energy which is efficiently
radiated away can be neglected by considering the long-term coupling
efficiencies of feedback energy to the ISM, and momentum injection can
be treated in a similar fashion.

Here, I first consider the injection of energy and derive the AGN
energy coupling efficiency which is required in order to explain the
observed break in the galaxy mass function (Section
\ref{sec:fagn_deriv}). Next, I use an analogous argument to derive the
required momentum loading factor (Section \ref{sec:fpagn_deriv}) and
present the expected variation of these factors with redshift (Section
\ref{sec:redshift}).

\subsection{Energy coupling} \label{sec:fagn_deriv}

Energy injection into the ISM comes primarily from two sources: stars
and active nuclei. The total amount of injected AGN energy, $E_{\rm
  AGN},$ integrated over the lifetime of the galaxy, depends on three
factors: the BHAR $\dot{M}_{\rm BH}$, the radiative efficiency
$\epsilon_{\rm AGN}$ of converting the mass into energy, and the
coupling efficiency $f_{\rm AGN}$ between the radiative energy output
and the ISM:
\begin{equation}\label{eq:agn_energy}
  E_{\rm AGN} = \int_0^{t_{\rm H}} f_{\rm AGN} \epsilon_{\rm AGN} \dot{M}_{\rm BH} c^2 {\rm d}t,
\end{equation}
where $t_{\rm H}$ is the Hubble time. Stellar feedback comprises
feedback at different stages in the star's life, but is also, on long
timescales, proportional to the total stellar mass formed:
\begin{equation}\label{eq:star_energy}
  E_* = \sum_i \int_0^{t_{\rm H}} f_{*,i} \varepsilon_{*,i} \dot{M}_* {\rm d}t,
\end{equation}
where the index of summation goes over the different feedback
processes, $\dot{M}_*$ is the star formation rate and
$\varepsilon_{*,i}$ is the energy release per unit mass formed for the
different processes. The product $\varepsilon_{*,i} \dot{M}_*$ then
gives the energy injection rate by stellar feedback of a particular
type. The ratio of energy input by AGN feedback to that of stellar
feedback is then
\begin{equation} \label{eq:ratio}
R \equiv \frac{f_{\rm AGN} L_{\rm AGN}}{\sum_i P_{*,i}} = \frac{f_{\rm AGN} \epsilon_{\rm AGN} c^2}{\sum_i f_{*,i} \varepsilon_{*,i}} \frac{\dot{M}_{\rm BH}}{\dot{M}_*},
\end{equation}
where the integration used in equations (\ref{eq:agn_energy}) and
(\ref{eq:star_energy}) can be removed since the limits are
identical. I also used $L_{\rm AGN} = \epsilon_{\rm AGN} \dot{M}_{\rm
  BH} c^2$ and $P_{*,i} = f_{*,i} \varepsilon_{*,i} \dot{M}_*$ for
conciseness. I will now derive the numerical value of $f_{\rm AGN}$
from the preceding equation, based on the following assumptions:
\begin{itemize}
\item Galaxies with $R < 1$ are dominated by stellar feedback, while
  galaxies with $R > 1$ are dominated by AGN feedback; therefore,
  there is a transition at $R = 1$, which is observable as the break
  in the galaxy luminosity function, the peak of the stellar-to-halo
  mass ratio, and the break in the black hole mass - velocity
  dispersion relation. A difference of the dominating feedback process
  is the conventional explanation of the shape of the galaxy stellar
  mass function \citep[see, e.g.,][]{Puchwein2013MNRAS,
    Moster2013MNRAS, Harrison2017NatAs}.
\item The radiative and feedback coupling efficiencies of AGN do not
  depend systematically on galaxy stellar mass. They might depend,
  directly or indirectly, on the gas content of the galaxy, but I
  argue below that the galaxy mass function must have been established
  at high redshift, when gas content was large in all galaxies.
\item The feedback energy per unit mass of stars formed, and the
  coupling of this energy to the ISM, do not depend systematically on
  galaxy mass. This is not necessarily completely true, but I show
  below that the possible dependence is smaller than the dependence of
  BHAR and SFR on galaxy mass.
\item Only the fraction of input energy which is efficiently coupled
  to the large-scale ISM is considered. This means that energy that is
  efficiently radiated away, as well as energy that is fully
  dissipated on small scales, is not taken into account.
\end{itemize}

With these assumptions, $R$ depends on galaxy mass only through the
BHAR-to-SFR ratio, while the other factors in eq. (\ref{eq:ratio}) are
all constant. There are various estimates for these constant factors
in the literature, which I outline below.

There are essentially two important modes of stellar feedback: winds
and supernovae \citep[radiation pressure is unlikely to be very
  important; see, e.g.,][]{Rosdahl2015MNRAS}. Stellar wind feedback
power, assuming a continuous star formation episode, is
\begin{equation}
P_{\rm w} \simeq 3\times 10^{41} f_{\rm w,p} \frac{\dot{M}_*}{\msun {\rm yr}^{-1}} {\rm erg s}^{-1},
\end{equation}
where $0.3 < f_{\rm w,p} < 1$ is an efficiency factor encompassing the
various uncertainties of wind production models
\citep{Leitherer1992ApJ}. The wind coupling efficiency $f_{\rm w,c}$
is less constrained, especially on scales larger than the star-forming
regions. Observational estimates suggest mechanical energy coupling on
star-forming region scale of $0.037 < f_{\rm w,c} < 0.38$
\citep{Rosen2014MNRAS}, while numerical simulations suggest a combined
kinetic and thermal energy retention of $0.23 < f_{\rm w,c} < 0.48$ at
the end of a star's wind phase \citep{Fierlinger2016MNRAS}. The
combined factor $f_{\rm w} \equiv f_{\rm w,p}f_{\rm w,c}$ can then
have values $0.011 < f_{\rm w} < 0.48$, however the lower end of this
range appears unlikely due to being estimated on smaller scales than
the whole galaxy. I will therefore adopt a range $0.2 < f_{\rm w} <
0.5$.

Supernova feedback can be estimated from either an energy argument or
a momentum argument. The energy estimate gives
\begin{equation} \label{eq:esn}
P_{\rm SN} \simeq 3\times 10^{41} f_{\rm SN} \frac{\dot{M}_*}{\msun {\rm yr}^{-1}} {\rm erg s}^{-1},
\end{equation}
where $f_{\rm SN} < 1$ is the very uncertain coupling efficiency, and
a \citet{Chabrier2003ApJ} mass function is used, giving a supernova
rate $\dot{N}_{\rm SN} = 0.01$ per Solar mass formed. A more robust
estimate can be attained by using the calculation in
\citet{Murray2005ApJ}, assuming momentum injection and taking the
terminal velocity of the wind to be approximately the velocity
dispersion in the galaxy:
\begin{equation} \label{eq:psn}
P_{\rm SN} \simeq \dot{p}_{\rm SN} \sigma_{\rm gal} \simeq 3\times 10^{40} \frac{\dot{M}_*}{\msun {\rm yr}^{-1}} v_{200} {\rm erg s}^{-1},
\end{equation}
where $v_{200}$ is the rotational velocity of the galaxy $v_{\rm rot}$
in units of $200$~km~s$^{-1}$ and I assume $\sigma_{\rm gal} = v_{\rm
  rot}/\sqrt{2}$. Combining this expression with eq. (\ref{eq:esn}), I
get $f_{\rm SN} \simeq 0.1 v_{200}$.  Numerical simulations of
individual and paired supernova explosions in various ISM geometries
give values $0.05 < f_{\rm SN} < 0.4$ \citep{Thornton1998ApJ,
  Walch2015MNRAS}. The highest efficiencies are reached on short
timescales or in simulations with multiple supernovae or in
pre-ionised clouds, so the average value should be smaller. In the
following, I will use a range $0.05 < f_{\rm SN} < 0.2$. This range
agrees well with the analytical estimate above, at least for galaxies
with $0.5 < v_{200} < 2$. Using the baryonic Tully-Fisher relation
\citep{Tully1977A&A}, this range of velocities corresponds to galaxy
stellar masses $5\times 10^9 \msun < M_* < 1.3 \times 10^{12} \msun$
\citep{McGaugh2005ApJ}, easily encompassing the allowed range of SMBH
scaling transition mass \citep{Martin2018arXiv} and the break in the
galaxy mass function \citep{Behroozi2013ApJ}. Therefore the possible
scaling of $f_{\rm SN}$ with galaxy mass via the terminal velocity of
SN-driven winds is unlikely to play a major role in determining the
transition mass between stellar- and AGN-feedback dominated galaxies.

The radiative power of luminous AGN is
\begin{equation}
L_{\rm AGN} = \epsilon_{\rm AGN} \dot{M}_{\rm BH} c^2 = 5.7 \times 10^{45} \epsilon_{0.1} \frac{\dot{M}_{\rm BH}}{\msun {\rm yr}^{-1}} {\rm erg s}^{-1},
\end{equation}
where I parameterize $\epsilon \equiv 0.1 \epsilon_{0.1}$. The
feedback power injected into the ISM is then $f_{\rm AGN} L_{\rm
  AGN}$.

Putting these expressions into eq. \ref{eq:ratio} gives
\begin{equation}
R \simeq 1.9 \times 10^{4} \frac{f_{\rm AGN}\epsilon_{0.1}}{f_{\rm w}+f_{\rm SN}} \frac{\dot{M}_{\rm BH}}{\dot{M}_*}.
\end{equation}
This equation can be rearranged to give $f_{\rm AGN}$:
\begin{equation}
f_{\rm AGN} \simeq 5.3 \times 10^{-5} R \frac{f_{\rm w}+f_{\rm SN}}{\epsilon_{0.1}} \frac{\dot{M}_*}{\dot{M}_{\rm BH}}.
\end{equation}

By definition, $R=1$ for galaxies where stellar and AGN feedback is
equally important. Using the transition mass $M_{\rm tr} \sim 3.4
\times 10^{10} \msun$ \citep{Martin2018arXiv} gives a BHAR/SFR ratio
$10^{-4} < \dot{M}_{\rm BH}/\dot{M}_* < 3\times 10^{-4}$, increasing
with redshift \citep{Yang2017ApJ, Yang2018MNRAS}. Expressing this as a
scaled relation $\dot{M}_{\rm BH} /\dot{M}_* = 3\times 10^{-4} f_{\rm
  a}$, with the scaling parameter $0.3 < f_{\rm a} < 1$, leads to
\begin{equation}\label{eq:fagn_final}
f_{\rm AGN} \simeq 0.18 \frac{f_{\rm w}+f_{\rm SN}}{\epsilon_{0.1} f_{\rm a}}.
\end{equation}

The full range of the adopted values of $f_{\rm w}$, $f_{\rm SN}$ and
$f_{\rm a}$ leads to a range for $f_{\rm AGN}$:
\begin{equation}\label{eq:fagn_num}
0.045 < f_{\rm AGN} < 0.42.
\end{equation}

\subsection{Momentum-loading factor} \label{sec:fpagn_deriv}

The above estimate has several inherent uncertainties, all related to
the coupling between injected energy and the ISM. An uncertain amount
of energy can be radiated away, and this fraction depends on the
details of the ISM. The interaction of stellar and AGN feedback can
also have unpredictable effect. It is impossible to eliminate these
uncertainties without detailed numerical simulations. On the other
hand, a similar estimate can be made by considering momentum injection
by both stars and AGN.

Stars mainly inject momentum via winds and supernova
explosions. Direct radiation pressure has a powerful, but mainly
local, effect in disrupting dense molecular clouds, but later rapidly
leaks out and does not couple very strongly to the galactic-scale ISM
\citep{Agertz2013ApJ}. Wind momentum injection is $p_{\rm w} \simeq
2.34 \times 10^{40}$~g~cm~s$^{-1}$ per Solar mass
\citep{Agertz2013ApJ}, which translates to
\begin{equation}
\dot{p}_{\rm w} = 7.43 \times 10^{32} f_{\rm p,w} \frac{\dot{M}_{\rm *}}{\msun {\rm yr}^{-1}} {\rm g \; cm \; s}^{-2},
\end{equation}
where $f_{\rm p,w} \simeq 1$ is the wind momentum loading factor. The
momentum injection rate from supernovae is
\begin{equation}
\dot{p}_{\rm SN} = 6.3 \times 10^{32} f_{\rm p,SN} \frac{\dot{M}_{\rm *}}{\msun {\rm yr}^{-1}} {\rm g \; cm \; s}^{-2},
\end{equation}
where $f_{\rm p,SN} \simeq 8-25$ is the supernova momentum-loading
factor \citep{Agertz2013ApJ, Martizzi2015MNRAS, Walch2015MNRAS}.

Using these values for stellar momentum injection, AGN momentum
injection $\dot{p}_{\rm AGN} = f_{\rm p,AGN} L_{\rm AGN}/c$ and the
fact that at transition mass, the two should be equal, leads to an
estimate for $f_{\rm p,AGN}$:
\begin{equation}
f_{\rm p,AGN} \simeq 3.3 \times 10^{-3} \frac{1.18f_{\rm p,w}+f_{\rm p,SN}}{\epsilon_{0.1}} \frac{\dot{M}_*}{\dot{M}_{\rm BH}},
\end{equation}
where the factor $1.18 \simeq 7.43/6.3$ is used to scale the wind
momentum injection rate to that of supernovae. Using the values of
$f_{\rm p,w}$, $f_{\rm p,SN}$ and the BHAR/SFR ratio leads to a range
of $f_{\rm p,AGN}$:
\begin{equation}\label{eq:pagn_num}
100 < f_{\rm p,AGN} < 830.
\end{equation}

\subsection{Redshift dependence} \label{sec:redshift}

\begin{figure}
  \centering
    \includegraphics[trim = 0 0 0 0, clip, width=0.48\textwidth]{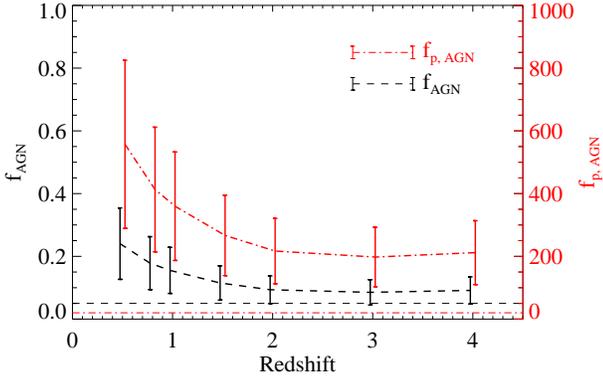}
  \caption{Required AGN energy coupling efficiency (black dashed line,
    left scale) and AGN momentum loading factor (red dash-dotted line,
    right scale) as function of redshift. Error bars represent the
    estimated range at each redshift, lines connect the mean values at
    each redshift. For clarity, the bars are offset horizontally by
    $\Delta z = \pm 0.025$. Horizontal lines represent analytical
    estimates of energy-driven AGN outflows: $f_{\rm AGN} = 0.05$ and
    $f_{\rm p, AGN} = 20$.}
  \label{fig:fagn}
\end{figure}

Both expressions for AGN coupling have redshift dependence which
arises from the variation of the BHAR/SFR ratio with redshift. At high
redshift, the ratio is typically larger, therefore the AGN coupling
efficiency or momentum loading factor can be smaller and still produce
the correct transition mass. Using the minimum and maximum BHAR/SFR
ratios at $M_* = M_{\rm tr}$ leads to the following results:
\begin{itemize}
\item At $z \simeq 0.5$, $0.13 < f_{\rm AGN} < 0.42$ and $290 < f_{\rm p, AGN} < 830$;
\item At $z \simeq 3$, $0.045 < f_{\rm AGN} < 0.13$ and $100 < f_{\rm p, AGN} < 290$;
\end{itemize}
Estimates of $f_{\rm AGN}$ and $f_{\rm p, AGN}$ for all redshift bins
considered in \citet{Yang2018MNRAS} are shown in Figure
\ref{fig:fagn}.

\section{Discussion}

The foremost implication of the derived range of $f_{\rm AGN}$ values
is that the coupling between AGN luminosity and the ISM should, on
average, be very large. Although I find $f_{\rm AGN} < 1$, which means
the estimate is possible on energy conservation grounds, the value is
generally larger than typical theoretical predictions and
significantly larger than observational estimates. Theoretical models
of AGN wind-driven outflow feedback \citep{King2010MNRASa,
  Zubovas2012ApJ} and radiation pressure feedback
\citep{Ishibashi2018MNRAS} predict $f_{\rm AGN} \sim 5\%$ under
idealised conditions, the same value is used in cosmological numerical
simulations \citep{Booth2009MNRAS, Schaye2015MNRAS}. Numerical
simulations of more realistic clumpy medium (Bourne \& Zubovas 2018,
submitted), as well as observations of real outflows
\citep{Fiore2017A&A} suggest even lower values $f_{\rm AGN} < 1\%$.

The required momentum loading factor shows much greater tension with
analytical predictions and observations. Models typically predict
$f_{\rm p, AGN} \sim 20$, a factor 5 lower than the lowest value
estimated here, although the scatter in analytical predictions is
larger than for $f_{\rm AGN}$ \citep{Zubovas2012ApJ}.

There are several potential ways to reconcile this tension. I first
consider the explanations based on assuming the estimated values of
$f_{\rm AGN}$ and $f_{\rm p, AGN}$ are approximately correct and later
discuss the possibility that they are significant and systematic
over-estimates of real values.

\subsection{Possibility of reaching high $f_{\rm AGN}$ and $f_{\rm p,AGN}$} \label{sec:phys_eff}

The derived coupling efficiency requires that a significant fraction,
much higher than $1\%$, of AGN luminous energy output is efficiently
coupled to the host galaxy ISM. The momentum loading factor similarly
suggests that the outflow should propagate in an ISM with very high
optical depth, which would lead to photons scattering many times
before escaping the galaxy, increasing the scalar momentum of the
outflowing gas. There are two ways of achieving this result.

One possibility is hyper-Eddington SMBH growth during Compton-thick
(heavily obscured) phases. In this case, most of the radiated energy
can couple to the ISM, producing warm absorbers \citep{King2014MNRAS,
  King2016MNRAS}. This energy can couple to the ISM surrounding the
AGN in multiple ways - via winds, radiation pressure, gas heating or
jets. The precise mechanism of this coupling is not important for the
purpose of this discussion, so long as it results in $f_{\rm AGN} \sim
1$ {\em during that stage of SMBH growth}. If heavily obscured
accretion results in the SMBH growing by $\Delta M_{\rm obs}$
throughout the lifetime of the galaxy, and the rest of the mass $M$ is
accumulated primarily during luminous accretion which has a low
feedback energy coupling efficiency $f_{\rm l}$, the average
coupling efficiency is
\begin{equation}
f_{\rm av} \sim \frac{\Delta M_{\rm obs} + f_{\rm l}\left(M-\Delta M_{\rm obs}\right)}{M} = f_{\rm l} + \left(1-f_{\rm l}\right) \frac{\Delta M_{\rm obs}}{M}.
\end{equation}
If, for example, $10\%$ of the SMBH mass is accumulated during heavily
obscured phases, and the rest produces energy-driven outflows with
$f_{\rm l} = 0.05$, we get $f_{\rm av} = 0.145$, an almost
threefold increase. Even if $f_{\rm l} \sim 0$, this results in
$f_{\rm av} = 0.1$, a value consistent with the derived prediction.

The fraction of mass that can be accreted during periods of strong
obscuration is limited by the feedback energy. If most of the radiated
energy is transferred to the ISM, it can be easily disrupted and
expelled far away. The AGN is then no longer obscured and $f_{\rm
  AGN}$ decreases significantly. \citet{King2016MNRAS} estimate that
the limiting mass for an AGN accreting in this hyper-Eddington mode is
decreased from the usual $M-\sigma$ value by a factor $\sim
\epsilon_{\rm AGN}^{1/2} \dot{m}^{-1/2}$, where $\dot{m}$ is accretion
rate in units of Eddington rate. The accretion rate must be $\dot{m}
\gg 1$ in order to produce strong obscuration, but cannot be too large
so that the limiting mass does not become too small. Therefore, I
predict that SMBHs grow $\sim 10\%$ of their mass in heavily obscured
phases with $\dot{m} \simgt 10$.

This prediction can be tested as more observations of obscured quasars
are made. Obscured growth should occur at high redshift, because
galaxies were more gas-rich then, making it easier to feed the SMBH at
hyper-Eddington rates. Also, at high redshift the required $f_{\rm
  AGN}$ and $f_{\rm p,AGN}$ are lower, therefore it is easier to
establish the SMBH-galaxy correlations. Both observations
\citep{Shields2006NewAR} and numerical simulations
\citep{Croton2006MNRASb} also suggest that black holes tend to grow
faster than their host galaxies, consistent with the result that a
lower $f_{\rm AGN}$ and $f_{\rm p,AGN}$ can establish the observed
galaxy mass function. Another prediction from this result is that the
break in the galaxy mass function and the SMBH transition mass were
established at $z \simgt 2$, consistent with observations
\citep{Behroozi2013ApJ}.

An alternative explanation for high $f_{\rm AGN}$ would be that most
AGN accretion occurs at very high radiative efficiency. This has two
helpful consequences: the required $f_{\rm AGN}$ decreases because it
is proportional to $\epsilon_{\rm AGN}^{-1}$, and the actual $f_{\rm
  AGN}$ increases, because at least in the wind-driven outflow model,
the fraction of energy transferred to the outflow is $\sim
\epsilon_{\rm AGN}/2$. If all SMBHs have maximal spins, $\epsilon_{\rm
  AGN} \sim 0.42$, the energy transferred to the outflow increases by
a factor $4.2$, while the required $f_{\rm AGN}$ decreases by the same
factor. This brings the two estimates into agreement.

There is some observational evidence that suggests typical values of
$\epsilon_{\rm AGN}$ to be higher than the $10\%$ assumed in the
calculations above \citep{Tombesi2010A&A, Tombesi2010ApJ,
  Tombesi2015Natur, Veilleux2017ApJ}. However, some theoretical
arguments suggest that the opposite should be true
\citep{King2005MNRAS, King2006MNRAS}. In addition, at least some of
the accretion on to SMBHs happens in radiatively inefficient modes, so
the total luminous output is less than the maximum possible. In
addition, observed large-scale outflows have $f_{\rm AGN} \ll 0.05$
\citep{Fiore2017A&A}, i.e. even if they are powered by rapidly
spinning SMBHs, the energy communication is much weaker than the
simple analytical estimate suggests. Therefore I think it is not very
likely that the discrepancy can be explained purely by invoking
high-spin SMBHs.

Finally, low-mass black holes can drive weak, but almost continuous,
outflows in gas-poor galaxies, provided that the outflowing gas
remains hot and approximately spherically symmetric
\citep{King2015ARA&A}. Although this process should only be important
at low redshift and in small galaxies, which are presumably below the
transition mass, it can nevertheless increase the total feedback
energy injected into the ISM over the lifetime of a galaxy.

Some of the arguments presented above can be used, qualitatively at
least, to explain the high values of $f_{\rm p,AGN}$ as well.
Momentum injection is also increased by obscuration, by an even larger
factor than energy injection, since high optical depth results in
multiple photon scatterings and a global increase in scalar
momentum. Furthermore, during highly obscured accretion, a large
fraction of injected energy is predominantly kinetic, coming from
hyper-Eddington winds, therefore the outflow momentum is also very
large. On the other hand, increased radiative efficiency does not
increase momentum injection, which is directly proportional to
luminosity.

\subsection{Selection effects}

Another important reason for the discrepancy might be a range of
selection effects that lead to observed $f_{\rm AGN}$ values being
much lower than real ones, whether momentary or long-term average.

The most important selection effect may simply be redshift dependence
of AGN outflows. As suggested above (Section \ref{sec:phys_eff}), the
connection between SMBH and the host galaxy may be established at high
redshift, before or simultaneously with the peak of star formation
rate density \citep{Madau2014ARA&A}. In this case, the outflows
observed in local galaxies are only much weaker analogues of outflows
that actually quenched star formation in their host galaxies. This
conclusion is supported by some observations which suggest that
quasar-mode feedback is inefficient in low-redshift AGN
\citep{Shangguan2018ApJ}. Current observations of massive outflows
\citep{Fiore2017A&A} do not show any redshift dependence of coupling
efficiency. However, molecular outflows, which seem to dominate the
mass and energy budget, have been detected only at low redshift
($z_{\rm mol} < 0.2$) so far. If molecular outflows are detected in
high-redshift galaxies, they should have higher energy coupling
efficiencies and momentum loading factors compared with those in the
local Universe.

It is also possible that each individual outflow goes through
(potentially multiple) stages of high and low coupling efficiency, but
outflows with low coupling efficiency are easier to observe. Low
coupling efficiency might be observed for several reasons:
\begin{itemize}
\item If the AGN has recently (less than $t_{\rm f} \sim r/v$ ago,
  where $r$ is the outflow radius and $v$ is its radial velocity)
  increased its luminosity, the outflow has not had time to react to
  this change and will be seen as inefficiently coupled. Such a
  situation may have occurred if an AGN that had faded recently was
  later reinvigorated by another feeding event. In outflow samples
  selected by AGN luminosity, this effect may be important. A
  recently-increased AGN luminosity also means the Eddington ratio is
  higher, therefore a negative correlation between outflow coupling
  and Eddington ratio is expected in this case. This phenomenon is
  unlikely to explain most of the discrepancy, because a significant
  fraction of observed AGN are accreting at low Eddington ratios
  \citep[and references therein]{Fiore2017A&A}. However, some
  individual sources may show such behaviour, potentially identifiable
  by having several spatially distinct outflows
  \citep{Nardini2018MNRAS}.
\item The outflow may have a higher coupling efficiency while its
  spatial extent is small ($r \lesssim 100$~pc), due to higher average
  gas density and lower incidence of possible low-density gaps through
  which most of the feedback energy might escape
  \citep{Nayakshin2012MNRASb, Zubovas2014MNRASb}. Outflows with low
  spatial extent are likely to be more obscured and therefore more
  difficult to detect, especially in gas-rich galaxies \citep[Section
    \ref{sec:phys_eff}; see also ][]{King2016MNRAS}. On the other
  hand, numerical simulations suggest the opposite, that the total
  kinetic energy rate of outflows increases with increasing radius,
  therefore this explanation is unlikely to be universal either.
\item A significant fraction of the outflow kinetic energy might be
  contained in material that is difficult to detect, for example due
  to low density, high ionisation, or rapid mixing with the galactic
  material making the outflowing material kinematically
  indistinguishable from undisturbed gas. This would mean that the
  observationally-derived coupling efficiencies are lower than real
  ones. However, non-molecular outflow components are observed to have
  much lower mass outflow rates, therefore their importance to the
  overall energy budget of the outflowing material is probably small
  \citep{Tadhunter2008MmSAI, Morganti2017arXiv}.
\end{itemize}

To summarize, it seems that selection effects related to individual
galaxies with outflows are unlikely to be able to explain the
discrepancy between observed and required values of $f_{\rm AGN}$.

\subsection{Uncertainties related to BHAR/SFR}

The derived value of $f_{\rm AGN}$ and $f_{\rm p, AGN}$ is inversely
proportional to the BHAR/SFR ratio, therefore uncertainties in that
value can have a significant effect on the results. The ratio derived
in \citet{Yang2017ApJ, Yang2018MNRAS} is an underestimate to some
extent, because broad-line and Compton-thick AGN were not included in
that work. Type 1 AGN comprise $\sim 20-30\%$ of all AGN
\citep{Lu2010MNRAS}, and the Compton-thick fraction is $\sim10-20\%$
\citep{Akylas2012A&A}, therefore a total of $\sim 30-50\%$ of AGN may
be unaccounted for and the average BHAR may be underestimated by as
much as a factor two. Correcting for this would lead to a reduction of
both $f_{\rm AGN}$ and $f_{\rm p,AGN}$ by the same factor, bringing
them closer to analytical predictions and observational constraints.

The correlation between BHAR and SFR has been the subject of much
debate. While numerous authors found an almost linear
\citep{Chen2013ApJ} or somewhat sublinear
\citep{Diamond-Stanic2012ApJ} relation, spatially resolved analysis
shows that the correlation is mainly observed on sub-kpc scales
\citep{LaMassa2013ApJ}. Large sample analyses \citep{Zheng2009ApJ}
suggest that the correlation is only present in statistically averaged
samples. Numerical simulations produce contrasting results, both in
favour of the existence of correlations in individual galaxies
\citep{Hopkins2010MNRASb} and against it, showing that the timescales
of SFR and BHAR changes are too different for meaningful correlations
to exist in individual galaxies \citep{Hickox2014ApJ,
  Thacker2014MNRAS, Volonteri2015MNRAS}. Overall, in individual
galaxies on short timescales, the dominant feedback mechanism may be
independent of galaxy mass. However, the argument presented in this
paper depends only on the long-term energy injection by AGN and
stellar feedback, therefore using long-term averages of BHAR and SFR
is appropriate. Individual galaxies should generally show some
correlation between long-term average BHAR and SFR values
\citep{Hopkins2010MNRASb, Thacker2014MNRAS}, therefore the argument
should hold on the scale of individual galaxies, and the uncertainty
in the transition mass is predominantly caused by other factors than
different timescale of BHAR and SFR variability.

The value of the transition mass between stellar- and AGN-dominated
galaxies determines the appropriate BHAR/SFR ratio. It is also quite
uncertain: \citet{Martin2018arXiv} give the mass as $M_{\rm tr} =
3.4\pm2.1 \times 10^{10} \msun$. Taking the values of $\dot{M}_{\rm
  BH}/\dot{M}_*$ corresponding to the lower bound of $M_{\rm tr}$
gives $0.093 < f_{\rm AGN} < 1.14$, while those corresponding to the
upper bound give $0.025 < f_{\rm AGN} < 0.44$. In either case, the
difference is less than a factor $2$ and is similar when individual
$\dot{M}_{\rm BH}/\dot{M}_*$ trends at each redshift are
considered. While the most optimistic scenario, i.e. a $z > 2$ galaxy
at the upper end of the allowed range for $M_{\rm tr}$, gives $f_{\rm
  AGN}$ consistent with predictions of theoretical models, the tension
with observed outflow properties remains unsolved. The variation of
$\dot{M}_{\rm BH}/\dot{M}_*$ around the mean value at a given redshift
is $\pm0.15$~dex for a galaxy sample at $0.5 \leq z < 1.3$
\citep{Yang2017ApJ}. This factor $\sim 1.4$ difference is lower than
that associated with uncertainty in transition mass.

\subsection{Uncertainties of stellar and AGN feedback efficiencies}

There are also significant uncertainties inherent in the evaluation of
the efficiency of stellar feedback \citep{Krumholz2014conf}. Feedback
processes not included in the consideration above, such as
photoionization feedback \citep[e.g.,][]{Krumholz2006ApJ,
  Goldbaum2011ApJ}, can control star formation in the host galaxy and
thus have a strong effect on the total efficiency of stellar feedback,
effectively adding another factor $f_{\rm ph} > 0$ to the term $f_{\rm
  SN} + f_{\rm w}$. Similarly, radiation pressure can add momentum to
the gas, adding a factor $f_{\rm rp} > 0$ to $f_{\rm p,SN}$ and
$f_{\rm p,w}$. This would increase the estimated $f_{\rm AGN}$ and
$f_{\rm p,AGN}$, leading to even stronger tension with
observations. On the other hand, geometrical effects of
multiple-source feedback \citep{Bourne2016MNRAS} can reduce stellar
feedback efficiency compared with single-source efficiency.

The fraction of massive star wind and supernova energy that is
injected as kinetic energy into the ISM is also quite
uncertain. Although the values used in the derivation here are based
on detailed numerical simulations, some authors have argued for much
lower actual coupling efficiencies, $f_{\rm SN} + f_{\rm w} < 0.1$
\citep{Fierlinger2016MNRAS}. These values are obtained from
simulations of long-term evolution of ISM disturbed by winds and
explosions of individual stars, although it is not straightforward to
extrapolate these results to stellar populations with multiple
feedback sources acting on long timescales. If $f_{\rm SN} + f_{\rm w}
= 0.1$ is used in eq. (\ref{eq:fagn_final}), the required AGN coupling
efficiency becomes $0.018 < f_{\rm AGN} < 0.06$, a range consistent
with analytical models of AGN wind feedback and cosmological
simulations, although still somewhat larger than values derived from
real observed outflows.

AGN feedback may also manifest in other forms than, or in addition to,
massive outflows, leading to higher actual values of $f_{\rm AGN}$
than outflow properties alone would suggest. Direct gas heating by the
radiation field can be important, although probably only in gas-poor
galaxies \citep{Sazonov2005MNRAS}. Radiation pressure on the ISM can
have significant effects \citep{Ishibashi2014MNRAS,
  Ishibashi2018MNRAS}, but these lead to outflows that should be
observable similarly to wind-driven ones. Jet feedback can be very
efficient compared to radiative output \citep{Heinz2007ApJ,
  Mezcua2014ApJ}, but it dominates only in low-luminosity sources
\citep{Heckman2014ARA&A}. In high-luminosity sources, such as
radio-loud QSOs, jet feedback can at most produce as much power as
radiative feedback, but typically has a radiative efficiency $3-5
\times 10^{-3}$ \citep{Merloni2008MNRAS}, i.e. is unlikely to increase
$f_{\rm AGN}$ significantly. Furthermore, jet feedback mostly affects
galaxies that have already ceased star formation, preventing
circumgalactic material from falling back in. Overall, it appears that
the energy input by AGN into the ISM of a gas-rich host galaxy is
dominated by massive molecular outflows.

Some of the factors determining $f_{\rm AGN}$ and $f_{\rm SN} + f_{\rm
  w}$ are the same, at least qualitatively, for both feedback
processes. For example, in a clumpy galaxy, more energy may `leak out'
to large distances and have little effect on the process of star
formation or SMBH feeding, independently of the origin of this energy
\citep{Zubovas2014MNRASb, Recchi2006A&A, Martizzi2015MNRAS}. The mass
of the galactic halo determines the conditions for outflow escape, and
can lead to collapse of outflow bubbles, also independently of their
origin. Therefore, conditions such as gas fraction or galaxy mass
should not have a direct effect on the ratio between AGN and stellar
feedback efficiencies; note that galaxy mass has a dominant effect on
the BHAR-SFR ratio, which is what determines the transition mass
between stellar- and AGN-dominated systems.

While the uncertainty of $f_{\rm AGN}$ and other coupling efficiencies
involved is currently large, over time it should decrease as more
detailed observations and numerical simulations are
performed. Eventually, this may enable the use of such coupling
efficiencies to investigate the variations in feedback efficiency
among individual galaxies or their populations selected by certain
parameters, such as stellar mass, morphology or environment. Assuming
that the individual coupling efficiencies are independent of galaxy
mass, one would be able to determine the relative importance of
stellar and AGN feedback processes in any given galaxy and see how
they correlate with other galaxy parameters. Alternatively, if the
ratio of stellar-to-AGN power input $R$ (eq. \ref{eq:ratio}) can be
determined for individual galaxies based on their mass compared to the
transition mass, coupling efficiencies can be derived for individual
galaxies, revealing the difference in how stellar and/or AGN feedback
operates in galaxies with different properties.

\subsection{Implications for individual galaxies across the mass range}

The primary result of this paper, namely the large required value of
$f_{\rm AGN}$ and $f_{\rm p,AGN}$, is derived considering galaxies
with $R=1$, i.e. those at the transition between stellar- and
AGN-dominated systems. Given the assumptions made in the derivation,
namely that all coupling efficiencies are independent of galaxy mass,
galaxies smaller (larger) than $M_{\rm tr}$ have $R<1$ ($R>1$) simply
due to the increase of the BHAR-to-SFR ratio with mass
\citep{Yang2018MNRAS}. However, this assumption is not necessarily
correct and individual galaxies may have AGN and/or stellar feedback
coupled more or less strongly to the ISM than the average efficiencies
described above. The reasons for these differences, and their
consequences, might be the following:

\begin{itemize}
\item A small galaxy may be AGN dominated if its $f_{\rm AGN}$ is
  particularly large, or stellar feedback is particularly
  inefficient. This may be the case in a post-starburst galaxy, where
  numerous supernovae have recently opened chimneys and other channels
  for energy to leak out from most of the galactic volume
  \citep{Recchi2006A&A}, but the central part is still gas-rich and
  feeds the AGN. Alternatively, a galaxy with low gas fraction, with
  low BHAR and SFR, may be more AGN-dominated if AGN feedback occurs
  via jets, which are more efficient than outflows in converting
  luminosity to mechanical energy \citep{Heinz2007ApJ,
    Mezcua2014ApJ}. However, as mentioned above, such galaxies should
  already have their stellar and BH masses established.
\item A large galaxy may be dominated by stellar feedback if its
  $f_{\rm AGN}$ is particularly small, or stellar feedback is
  particularly efficient. This may be the case if the AGN is fed
  slowly enough that a radiatively inefficient accretion flow develops
  \citep{Yuan2014ARA&A}.
\item A major gas-rich galaxy merger leads to a starburst, followed by
  a period of AGN activity after several times $10^8$~yr. AGN feedback
  may then quench star formation or even enhance it
  \citep{King2015ARA&A}. In addition, the changes in gas morphology
  due to the merger and its associated feedback can lead to wild
  variations in feedback coupling efficiencies. However, this effect
  should not be dominant to the establishment of galaxy-scale
  correlations, because mergers only contribute a small fraction to
  the total star formation and luminosity of galaxy population at any
  redshift \citep{Hopkins2010MNRASc}.
  
\end{itemize}

All of these effects would lead to the transition between the two
feedback regimes becoming more blurred. The large scatter of black
hole masses around the transition region suggests that this is
probably the case. Detecting individual galaxies where AGN or stellar
feedback efficiencies are very different from the mean values would be
possible with large data sets.

\section{Summary and conclusions} \label{sec:concl}

In this paper, I presented an empirical estimate of the AGN feedback
coupling efficiency $f_{\rm AGN}$, i.e. the fraction of AGN luminous
energy output that is injected into the host galaxy ISM. This estimate
relies only on two assumptions: that the total energy injected by AGN
and by stellar feedback is the same for galaxies with stellar mass
equal to the `break', or transition, mass between
stellar-feedback-dominated and AGN-dominated galaxies; and that the
coupling efficiencies for AGN and stellar feedback processes are
independent of galaxy mass. The estimated efficiency is, on average,
very large, $f_{\rm AGN} > 0.045$, and remains $> 0.01$ even accounting
for possible systematic uncertainties. Such high efficiency can only
be achieved in very highly obscured AGN during the warm absorber
phase, and perhaps in energy- or radiation-pressure-driven large-scale
outflows. However, observed outflows typically have lower coupling
efficiencies. A similar tension is seen when momentum loading is
considered instead of energy injection: the required value $f_{\rm
  p,AGN} > 100$, much higher than analytical or observational
estimates.

There are several possible ways of resolving this tension:
\begin{itemize}
\item A significant fraction of total AGN feedback energy may be
  injected into the ISM during heavily obscured phases, when most of
  the luminous energy is used for driving the gas;
\item Selection effects may result in only inefficient outflows being
  observed, although this appears unlikely;
\item AGN-induced quenching of star formation may have happened
  predominantly at high redshift ($z > 2$), and outflows observed in
  the local Universe are quantitatively very different from those that
  actually quenched their galaxies;
\item Stellar feedback coupling efficiencies might be significantly
  lower than those used in calculating this estimate.
\end{itemize}
None of these possibilities is mutually exclusive with any other.

As observational data become better and detailed numerical simulations
provide ever more information on the relevant coupling efficiencies,
an estimate of $f_{\rm AGN}$ such as provided in this paper may be
used to test various feedback models and to analyse variations in
feedback among different galaxy sub-populations.

\section*{Acknowledgments}

I thank Sergei Nayakshin, Andrew King, Mar Mezcua and Guang Yang for
helpful suggestions during the preparation of this manuscript. I also
thank the anonymous referee for insightful comments that helped
improve the clarity of the arguments presented here. This research was
funded by a grant (No. LAT-09/2016) from the Research Council of
Lithuania.


\end{document}